# Vertical-oriented 5G platform-as-a-service: user-generated content case study

Sarang Kahvazadeh, Hamzeh Khalili, Rasoul Nikbakht Silab, Bahador Bakhshi, Josep Mangues-Bafalluy
Centre Tecnològic de Telecomunicacions de Catalunya (CTTC/CERCA)
Castelldefels, Spain
Emails:{ skahvazadeh, hkhalili, rnikbakht,bbakhshi, jmangues }@cttc.es

*Abstract*—**5G realizes an impactful convergence, where Network Functions Virtualization (NFV) and cloud-native models become fundamental for profiting from the unprecedented capacity offered at the 5G Radio Access Network (RAN). For providing scalability and automation management over resources in 5G infrastructure, cloud-native and Platform as a service (PaaS) are proposed as solutions for paving the way for vertical applications in 5G. This paper leverages cloud-native models, PaaS, and virtual testbed instances to provide key platform provisioning and service life-cycle management features to a selected User Generated Content (UGC) scenario in multimedia applications. Specifically, this article and results show how service-level telemetry from a UGC cloud-native application is used to automatically scale system resources across the NFV infrastructure.**

## I. INTRODUCTION

Nowadays multimedia streaming is one of the most popular services over the Internet. The disruption of cloud/edge computing and internet of things (IoT) into 5G architecture poses virtualization (i.e., NFV) as a fundamental tool for supporting massive UGC multimedia streaming. Such disruption heavily influences multimedia design; this being dramatically changed by container-based micro-services [1]. That is, the monolithic approach (all application functions in a single resource) are replaced by a distributed view, in which the application is split into micro-services hosted inside lightweight virtualization containers. This is referred to as cloud-native computing in software development. A 5G multimedia application is then expected to be composed of several containers across several physical nodes at the cloud/edge of the communications infrastructure, leveraging network overlays and specific reference points for communication and ultimately providing a service.

One of the main concerns with cloud-native 5G multimedia application relates to scalability. Scalability is about remaining unchanged the service-level performance indicators despite demand. Scalability may also have to conform with other criteria, like cost savings (e.g., Mobile Network Operator may impose additional charges for resources consumed at the edge of the network). Container Orchestration Engines (COE) (Kubernetes [2] considered the de facto standard) are the common choice for deploying cloud-native applications at the resource-constrained edge of the 5G network. Kubernetes enables cloud-native life cycle management such as self-healing, automatic scheduling, automatic scaling, application management, creating predictable infrastructure, etc.

Kubernetes facilitates the scaling of resources according to system telemetry via its Horizontal Pod Autoscaler (HPA) [3] controller. Traditionally, HPA only was setup for specific metrics such as CPU and Memory data for conditioning the scaling out/in of Pods (Pods are Kubernetes's smallest deployable computing unit. A Pod may contain one or more containers). Nevertheless, HPA may also be configured to feed from custom service-level metrics from Pods, other Kubernetes Objects, or even external sources. This ability yields freedom to design automatic scaling algorithms subject to arbitrary service-level criteria.

In order to Kubernetes hosts micro-services for handling HPA custom metrics, we also need to have proper 5G infrastructure and testbed. To be specific for handling scalability issue in vertical and horizontal way, we need also to offer a scalable 5G testbed environment where vertical applications such as multimedia streaming can easily be tested with 5G infrastructure alongside with Kubernetes. This scalable testbed can offer testing as a service. In this paper, we illustrate our testbed architecture and design that can be scaled to different instances for easing vertical´s applications to be tested with 5G infrastructure for different scenarios and use-cases. Our testbed also offers a novel Platform as a Service (PaaS) where Kubernetes infrastructure are implemented as network slices sub instances (NSSI). The PaaS is designed and implemented aligned with ETSI NFV IFA 029 [4]. We also show that how we can get benefits over our infrastructure in the case of scalability. Finally, in this paper, we present this concept through a chosen case study scenario that is the emulation of a crowded UGC (e.g., live sports or news coverage) where caching/processing applications endpoint hosted at Multi-access Edge Computing (MEC) are dynamically scaled by HPA conditioned to a custom service-level metric.

## II. Related Work

Recently, there has been a lot of activity in the development of 5G testbeds for supporting research and development in mobile technologies [5]. In this section, our focus is placed on the cloud-native models or virtual testbed instances, to provide key infrastructure provisioning and service life-cycle management features, supported by a variety of testbed providers in EU projects. For this, we summarized a few of them here. One of the relevant 5G testbed is TNOR's testbed facility that has been leveraged from 5G-VINNI project [6]. This testbed is proposed to support end-to-end network slicing, service deployment and management, toward onboarding and running experiments and validation of relevant use cases in 5G mobile network. In addition, The NFVI and VIM are based on OPNFV, with OpenStack as VIM and KVM as hypervisor. The NFVO provides network service orchestrator and resource orchestrator functions. The 5G-Berlin testbed [7] provides a 5G standalone network to test 5G technologies like NFVO, function virtualization, SDN, open-source stack, mobile edge computing and network slicing for the implementation of applications for verticals in mobile networks. 5GENESIS testbed is leveraged from 5GENESIS project [8], providing heterogeneous physical and virtual network elements under a common openness framework for experimenters and vertical industries. The testbed enables end-to-end slicing and cloud-native NFV MANO capabilities for deploying and running variety of experimenter applications (NetApps) on top of NFV infrastructure.

Nevertheless, despite all functionalities offered by other testbeds, cloud-native model and Platform as a Service are not provided together as an underlying 5G infrastructure. The combination of them allowing 1) vertical applications, 5G internal services and 5G Core functions to share NFVI resources, and 2) experimenters to have a simplified and sufficient abstracted visibility.

## III. Virtual TestBed Description

In this section, we propose and illustrate our testbed instances together with PaaS design and architecture over our 5G infrastructure. These testbed instances alongside with PaaS bring scalable 5G infrastructure and scalable testing as a service for application's owners.

### A. Testbed Infrastructure

Our testbed is based on Amarisoft 5G RAN, Open5GS core, and ETSI OSM MANO stack. It leverages a Cloud Radio Access Network (C-RAN) architecture with virtualized 5G core alongside with 5G RAN.

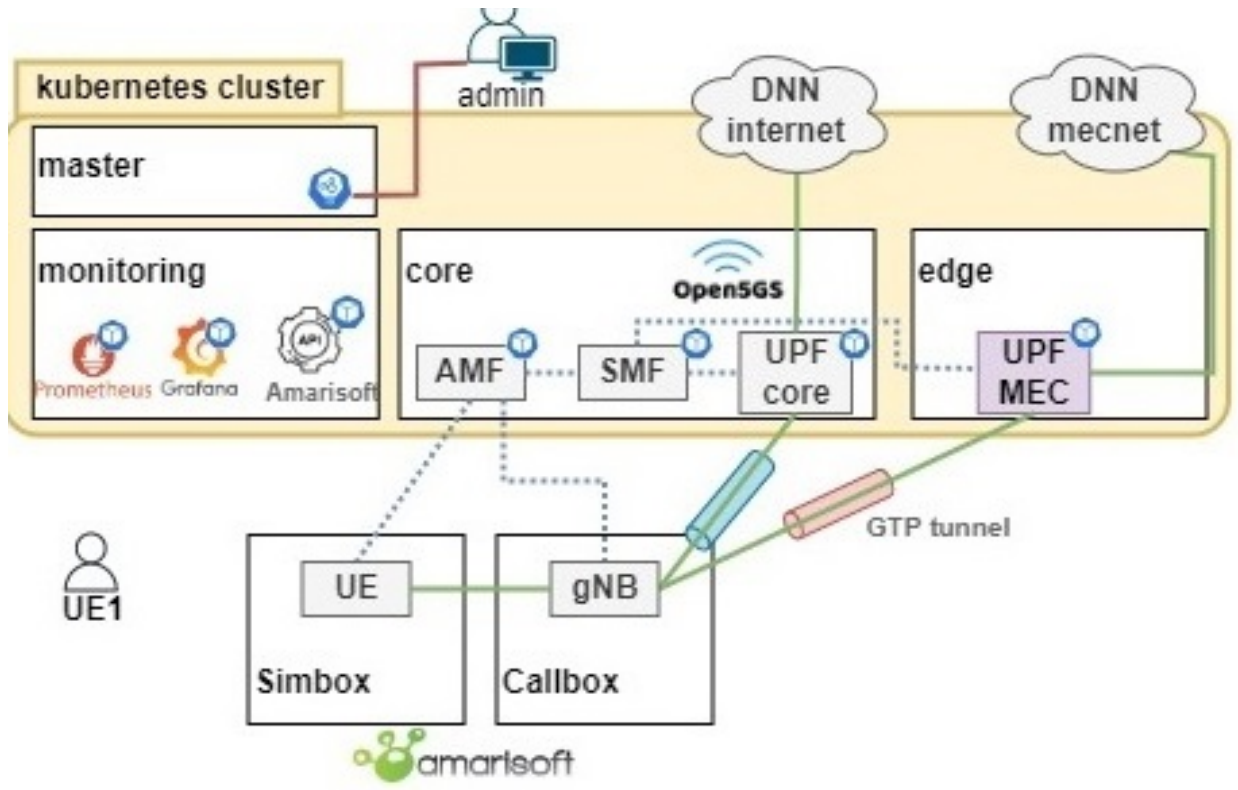

Figure 1. Amarisoft RAN/Open5gs Core (AMF: Access and Mobility Management Function, SMF: Session Management Function, UPF: User Plane Function, DNN: Data Network Name, gNB: Next Generation Node B, UE: User Equipment).

5G RAN-Core segment is implemented with the industry Amarisoft 5G Private Node, which provides 5G RAN capabilities as well as management northbound interfaces (NBI) out-of-the-box, enabling the establishment of 5G end-to-end services. The Amarisoft Northbound Interface (NBI) exposes a range of telemetry and actuation via web sockets, which may be leveraged for emulating a wide variety of scenarios and management algorithms. Also, the 5G core in our testbed is Open5GS, that is implemented in Kubernetes infrastructure and connected to our Amarisoft Radio Access Network (RAN). Figure 1 shows an implemented example of 5G RAN/CORE. Edge and Cloud 5G segments are enabled thanks to Virtualized Infrastructure Managers (VIM) such as OpenStack and Kubernetes. The former is typically used to provide Infrastructure as a Service (IaaS), while the latter helps deploy Mobile Edge Computing (MEC) architecture, state-of-the-art cloud-native applications, and dynamic infrastructure/service support components (aligned with ETSI NFV IFA 029).

### B. Testbed Instances

Our testbed also provides a sandbox for creating isolated Cloud/MEC environments. These are referred to as Testbed Instances (TI). A Testbed Instance will be deployed for testing and validation of our scenarios. Figure 2 shows the overall view of default resources of our TIs.

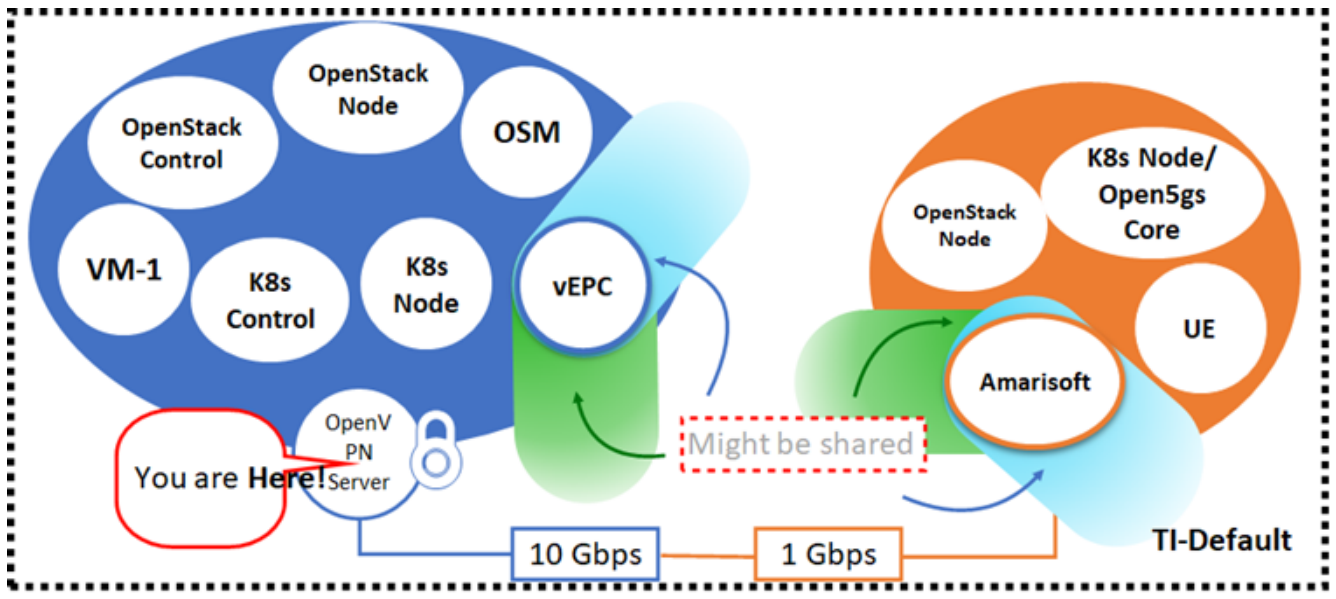

Figure 2. TI's default resources.

A default testbed instance consists of the following resources:

- Cloud Resources: High speed underlay, Network Function Virtualization Orchestrator (NFVO), Virtual Infrastructure Manager (VIM) control, Kubernetes control and Virtual Machines (VMs).
- Edge/RAN Resources: Lower speed underlay, Amarisoft 5G Private Node, Open5GS core in Kubernetes (K8s), VIM compute, K8s compute and User Equipment (UE) instance.

TI's Resources are illustrated in Figure 3 below.

C. Internal testbed instances design

The internal testbed instances include two levels: 1) the individual 5G testbed infrastructure elements and 2) monitoring modules for the real-time metrics recording.
Infrastructure level details:
The testbed benefits from Sandbox mechanisms for creating/building a testbed instance environment and design and creating Testbed Instance Template (TIT) composed of multiple stateful LXD VMs, which may contain:

a. NFVI: OpenStack Cluster, Kubernetes Cluster,
b. Virtual Machines (VMs),
c. ETSI OSM (whatever release),
d. Amarisoft RAN along side with Open5GS core implemented in Kubernetes.

Note that the RAN is shared among all created TIs, thus access to the RAN should be coordinated accordingly.
Metrics level:

- Our platform provides NetData for gathering Virtual Network Function (VNF) or resource telemetry, also, as upstream telemetry aggregator, e.g., slice-level.
- Prometheus can be configured for long-term metrics storage.

The figure below illustrates TI's architecture and template examples.

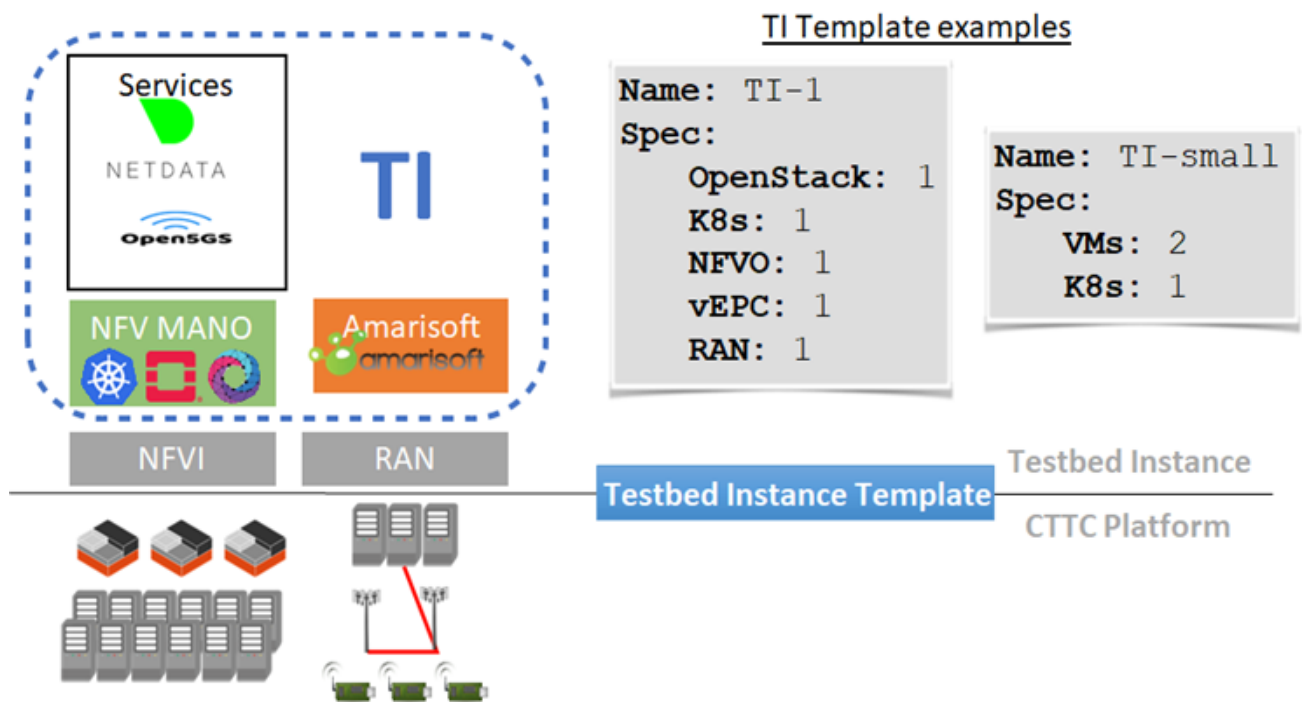

Figure 3. TI components, architecture, and TIT examples.

D. Technologies and protocols in testbed instances

Tools involved in TI implementation are as following:

1. Service management tools: Version control (Gitlab), PaaS (Kubernetes v1.18), PaaS Controller (Kubernetes Federation API),
2. Edge and cloud infrastructure: VIM (OpenStack),
3. ETSI NFV MANO: OSM v11,
4. GPU: NVidia RTX 2080Ti,
5. RAN Controller.

Finally, Figure.4 shows a created testbed instance.

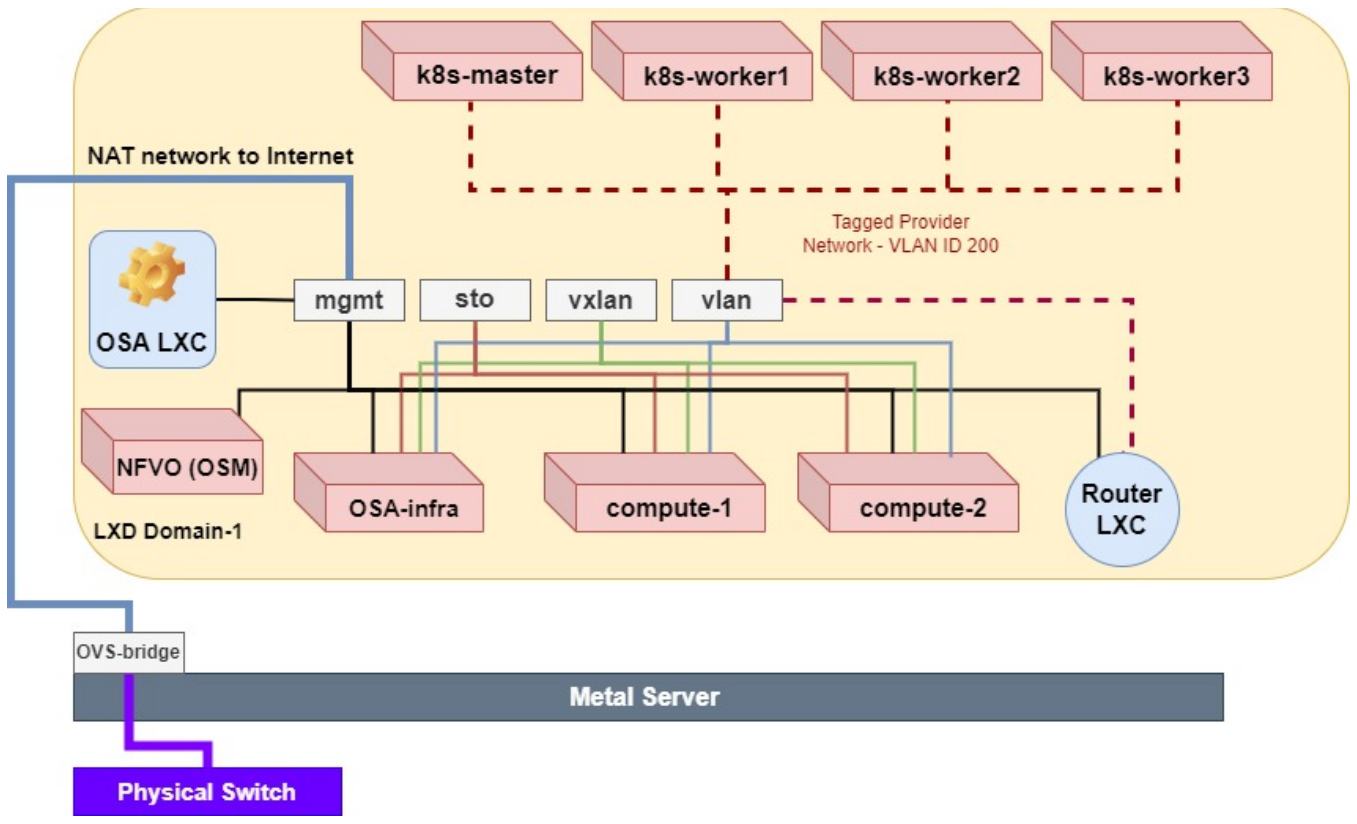

Figure 4. Testbed Instance alongside PaaS.

E. Platform as a Service (PaaS) implementation

The designed PaaS follows "cloud-native" design principles. The PaaS abstraction layer particularly facilitates the utilization of container technologies, where the consumer does not manage or control the underlying cloud infrastructure including network, servers, operating systems, storage and platform services, but has control over the deployed applications and possibly over application hosting environment configurations. The concept of PaaS been introduced in ETSI GR NFV-IFA 029 [4].

In this work, we focused on the use case where VNFCs (Virtual Network Function Containers) are deployed in containers in virtual machines. The virtualization layer in NFVI is responsible to provide: 1) the VM runtime environment, 2) the OS in the VM, 3) the container runtime environment for the containerized VNFCs hosted by the VMs running on a hypervisor layer of the NFVI-nodes and 4) the virtual network so that the containers can communicate within a VM, an NFVI-Node and across NFVI-Nodes. The NFV-MANO allocates the resources needed by the VNFCs. A "Container Infrastructure Service Management" function will provision the containers, when the VNFCs are deployed. It can be a separate functional block or integrated in an NFV-MANO component. VNFCs, which are containers, are part of the VNF (while their runtime environment is provided by the NFVI).

Following the ETSI GR NFV-IFA 029 specifications, we are adopting our design with the option of deploying container-based services on top of a shared NFVI. The PaaS-inspired design with the capability of supporting container-based services is composed of the Container Infrastructure Service Instance (CISI) and Container Infrastructure Service Management (CISM). CISI is the execution environment for the container cluster where the container-based services run CISM is responsible for the infrastructure resource management and lifecycle management of the execution environment for container cluster. CISI and CISM realize the PaaS instance. In our approach CISI and CISM are part of the VNF. The CISM is responsible for managing the lifecycle of CISI and initiating container-specific virtual resource management based on the virtual machines allocated to the VNF. The VNF is implemented as VNFCs whereby each VNFC is deployed as a container in a virtual machine (see Figure.5, Figure.6).

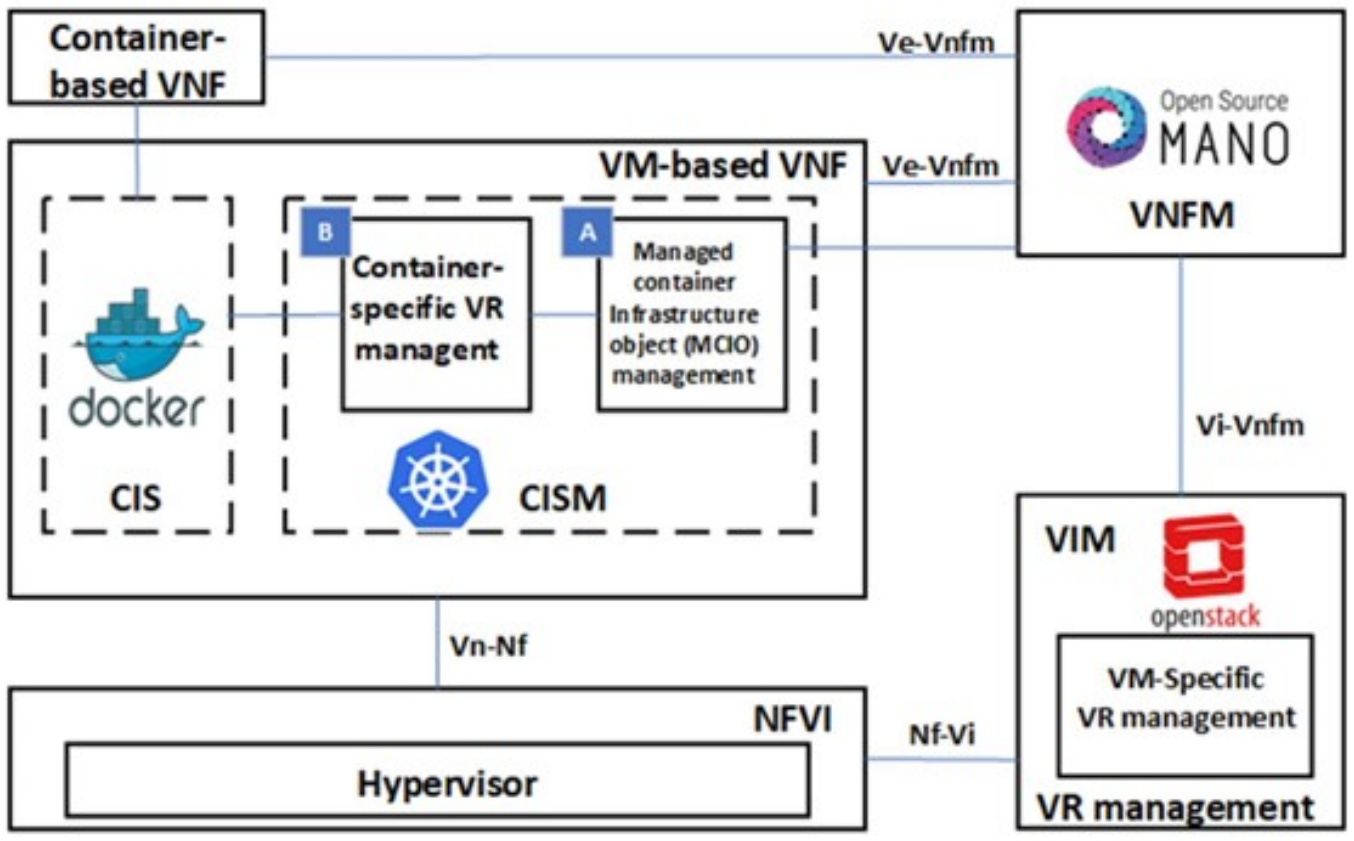

Figure 5. CISM embedded into VNF with support for shared container service.

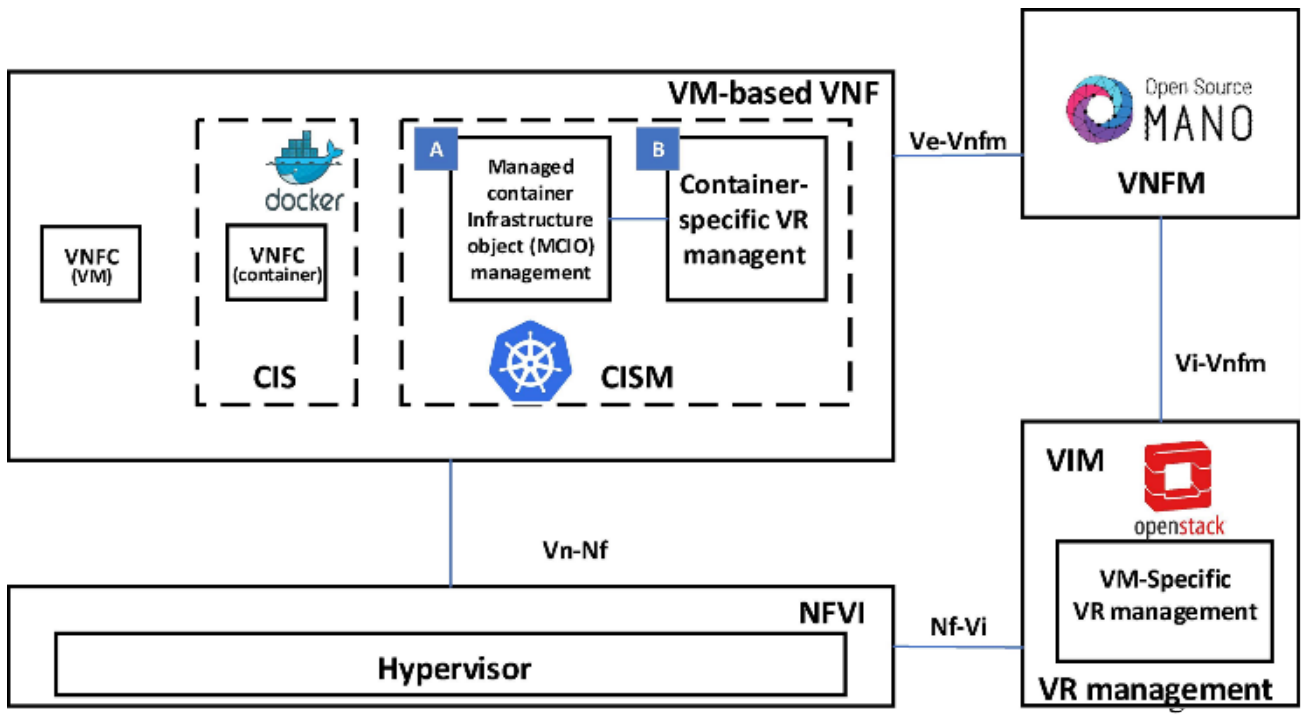

Figure 6. CISM embedded into VNF without support for shared container service.

In our design, resource orchestration is performed by an ETSI Open-Source MANO (OSM) NFVO. OSM supports descriptor files written in yaml (Virtual Network Function Descriptors - VNFDs and Network Service Descriptors NSDs). The VNFD defines the needed VNF resources in terms of compute resources and logical network connection points, the image that will be launched at the VM, as well as the autoscale thresholds based on the metrics that are being collected from the Telemetry service of the VIM. The NSD is responsible for the connection point links, using virtual links, among the interconnected VNFs, mapping them to the physical networks provided by the VIM. The primary aim of the OSM MANO is to coordinate NFV Infrastructure (NFVI) resources and map them efficiently to various VNFs. In turn, VNFs can then be interconnected into chains to realize more complex Network Services (NS). OSM reserves a number of VMs on top of which the Kubernetes clusters will be deployed. Thus, Kubernetes will be deployed as overlay. The underlay is the respective VIM (i.e., Openstack) and OSM. Part of the resource orchestration is also the scale-out and scale-in. The NFV MANO is able to scale dynamically resources to support the VNFs according to the heterogeneous quality of service requirements of the applications at hand. That is, Key Performance Indicators (KPI) of the VNF are monitored and if they are above a given threshold a scaling process is started, which implies the creation of new VMs to deploy the VNF (or to accommodate the Kubernetes clusters).

After describing our testbed instances and PaaS designs and implementation, a high-level figure for our testbed for testing the following scenario is presented in Figure.7. Technologies that are used for the implementation of PaaS are Kubespray [9], sets of ansible servers for automation [10] and OSM DAY0 cloudinit files [11].

When applications are deployed in PaaS (Kubernetes) to be tested with 5G, we are able to scale applications with HPA inside of Kubernetes infrastructure and/or even scaling by creating more PaaS´s NSSIs.

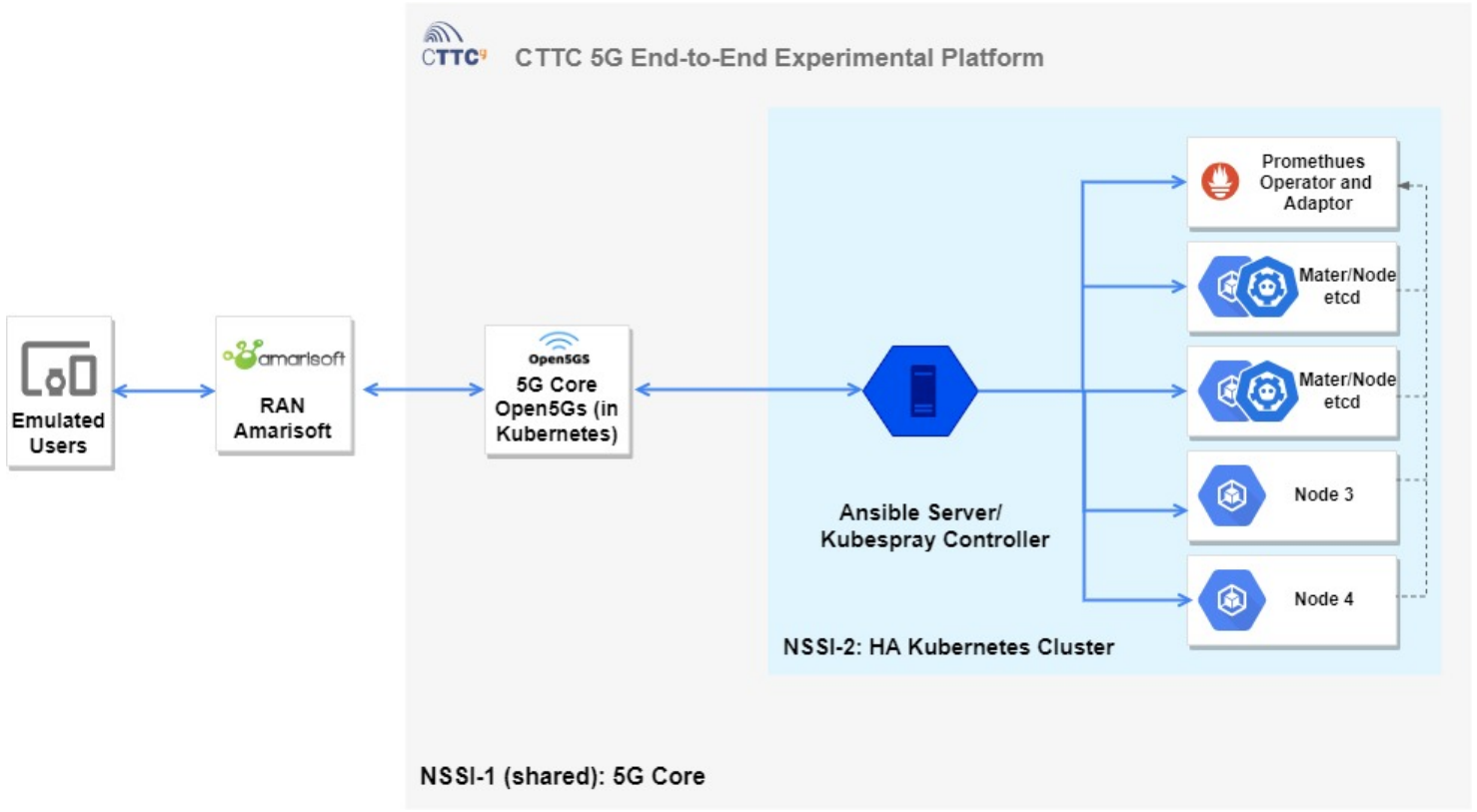

Figure 7. Testbed high-level architecture.

## III. Multimedia Scenario Description

Focusing on Multimedia domain, High Quality User Generated content (UGC) production services uploading in a dense area is one of the most impacting 5G UCs. In this scenario the UGC in the form of 4K video will be produced with smart phones and streamed up-link via 5G to a cloud-based system where it will be integrated with a professional video production. It is expected that 5G supports very high-density of devices, as well as high concurrent up-link transmissions capacity from one single location, avoiding congestion far beyond what is possible by existing 4G networks. Specifically, we demonstrate a Network Slice Subnet Instance (NSSI) holding a tenant service able to balance the load of incoming UGC streams coming from RAN. Such NSSI is orchestrated with ETSI OSM as NFV Orchestrator and constitutes the Virtual Network Function (VNF) resources on top of which the tenant service (i.e., COE and cloud-native application) is executed.

Focusing on the tenant service (i.e., cloud-native 5G multimedia application), it leverages IPVS as load balancer for a dynamic pool of endpoints, i.e., Video Sink Pods (VSP). As the configured HPA metrics change, the tenant service scales-out/in the number of VSP, therefore dynamically changing the total capacity of supported UGC streams and resources footprint. That is, allocating more resources when UGC streams increase, while destroying unused resources when no longer needed.

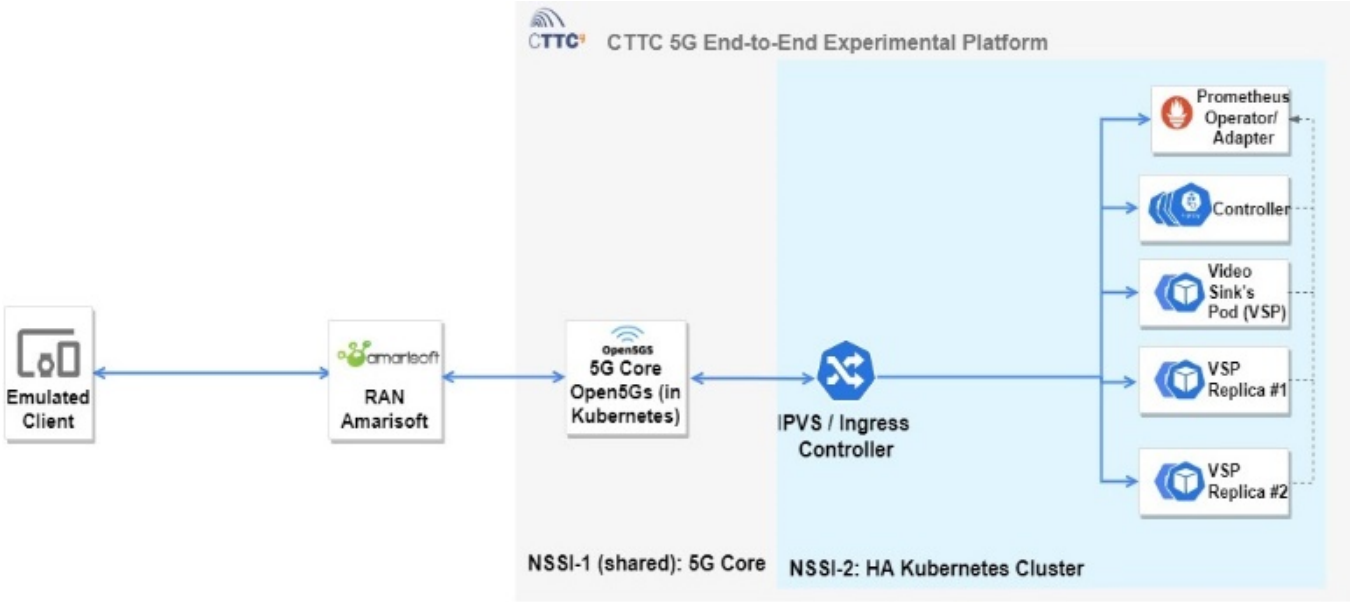

Figure 8. high-level deployment of multimedia architecture.

Figure.8 shows a simplified view of the proposed scenario. NSSI-2 shares access to the 5G Core. NSSI-1 is a shared Open5GS core implemented in Kubernetes where UPF can be at the edge or core according to Figure. 1. That is, UGC streams will come out of 5G Core and leverage NFV networking to reach NSSI-2. The scenario is implemented over our testbed instance that leverages PaaS where Kubernetes infrastructure is implemented as NSSI. At arrival, streams are balanced by IPVS to available VSPs. The latter being dynamically added or removed by HPA according to custom monitoring data deposited at a local Prometheus instance by the tenant service. Specifically, a new Video Sink Pod (VSP) instance is set to be automatically created if the average active streams per VSP surpasses a determined threshold.

## IV. TEST DESCRIPTION AND WORKFLOW

The development of this scenario will allow testing different combinations of load balancing algorithms and various thresholds determining scaling out of VSP. Figure.9 shows the workflow from an experiment where Emulated Clients acting as UGC requests to upstream a+b+c+…+z streams, but only a+b+c streams are initially admitted due to limited VSP http_requests. Then, HPA operations extend the pool of VSP, which in turn admit the remaining d+e+…+z UGC streams. The following illustrates the experimental workflow:

1. Emulated Clients send request to the controller.
2. Emulated Clients get notified of a VSP endpoint from controller.
3. HPA is configured to scale-out VSP according to a custom metric, http_requests. Specifically, a new VSP is created if it receives more than s=1000 requests.
4. Emulated Clients request the upload of (a+b+c+...+z) UGC streams to the VSP endpoint.
5. IPVS load balances the upstream a+b+c+...+z to an existing Video Sink Pod (VSP1).
6. VSP1 accepts upstreams a+b+c, but upstreams d+e+…+z are denied due to an overloaded VSP1. The rest of streams are denied.
7. HPA notices the observed metric surpassing the predefined threshold and automatically creates an additional VSP. IPVS then gets informed of the newly created VSPs and can now balance traffic to it.

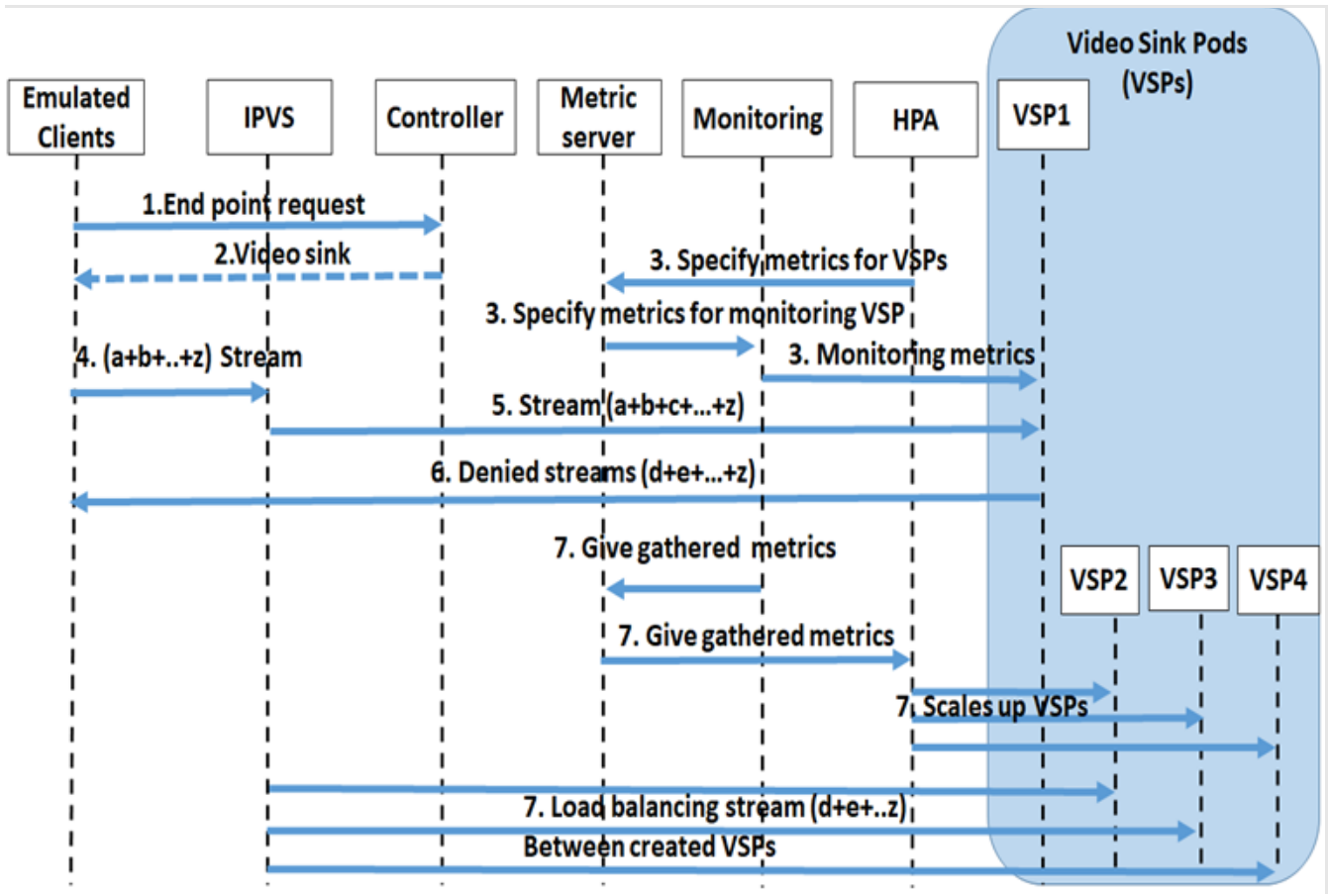

Figure 9. Multimedia scenario workflow.

VPSs leverage prometheus-fastapi-instrumentator [12] and starlette-exporter [13] Python libraries for exposing service-level metrics to Kubernetes. This way HPA may condition scaling operations on them. Figure.10 illustrates a single VSP and HPA before Emulated Clients start requests. Client request is simulated with Locust python library [14] with 100 users with 10 users spawned/second Hatch rate.

```
NAME                      READY   STATUS    RESTARTS
vsp-7c64fd69cf-qsw9j      1/1     Running   2
NAME    REFERENCE          TARGETS   MINPODS   MAXPODS   REPLICAS
vsp     Deployment/vsp     33m/1     1         10        1
```

Figure 10. VSP and HPA creation.

When Emulated Clients start sending requests for VSP, the associated http_requests increase. This is when HPA detects a threshold violation in the observed custom metric and starts scaling out VSP. This is illustrated in Figure.11, specifically when http_requests > 1 (1000m where m stands for milli-units, or 1000ths of a unit) such as 2913m, 1946m, and 1433m, HPA starts to create VSP replicas until http_requests become 942m, which is less than the set target value, 1 (1000m).

```
NAME    REFERENCE          TARGETS   MINPODS   MAXPODS   REPLICAS
vsp     Deployment/vsp     2913m/1   1         10        3

NAME    REFERENCE          TARGETS   MINPODS   MAXPODS   REPLICAS
vsp     Deployment/vsp     1946m/1   1         10        5

NAME    REFERENCE          TARGETS   MINPODS   MAXPODS   REPLICAS
vsp     Deployment/vsp     1433m/1   1         10        9

NAME    REFERENCE          TARGETS   MINPODS   MAXPODS   REPLICAS
vsp     Deployment/vsp     942m/1    1         10        10
```

Figure 11. Lifecycle of HPA user requests.

Figure.12 illustrates created VSP´s replicas.

```
vsp-7c64fd69cf-4wmsr      1/1     Running   0
vsp-7c64fd69cf-cs6jl      1/1     Running   0
vsp-7c64fd69cf-csx47      1/1     Running   0
vsp-7c64fd69cf-hm6fl      1/1     Running   0
vsp-7c64fd69cf-j8rvx      1/1     Running   0
vsp-7c64fd69cf-jl6dl      1/1     Running   0
vsp-7c64fd69cf-qsw9j      1/1     Running   2
vsp-7c64fd69cf-rfnsc      1/1     Running   0
vsp-7c64fd69cf-wkp5t      1/1     Running   0
vsp-7c64fd69cf-zv78c      1/1     Running   0
```

Figure 12. Created VSP´s replicas.

After the required VSP replicas are created, IPVS controls traffic distribution among them. Figure.13 shows the distribution of traffic among 5 of VSP replicas. Consider in IPVS, you have a choice for selecting different algorithms for traffic distribution among created VSP´s replicas

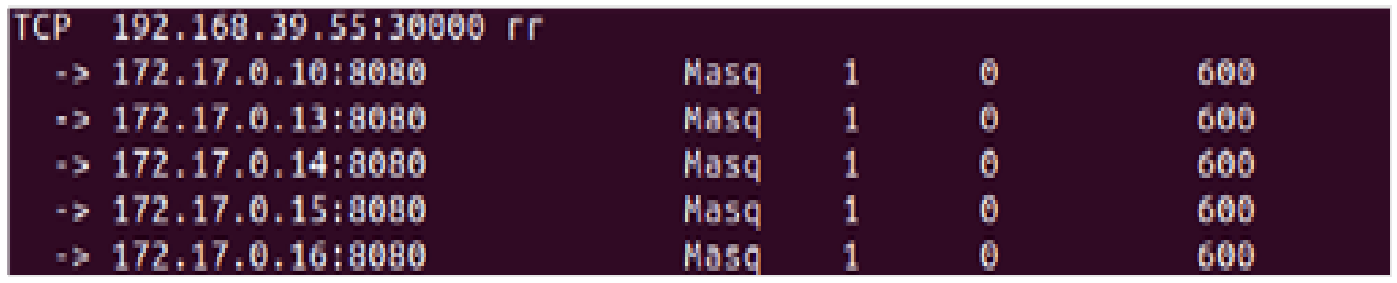

Figure 13. IPVS load balancing.

Figure.14 shows all the HPA process during our test, such as scaling out/in of VSP. The blue and red lines are min and max number of replicas, set to 1 and 10, respectively. The yellow line shows the running replicas. As we can see, HPA automatically creates enough VSP replicas during the test, and removes them after the requests finish.

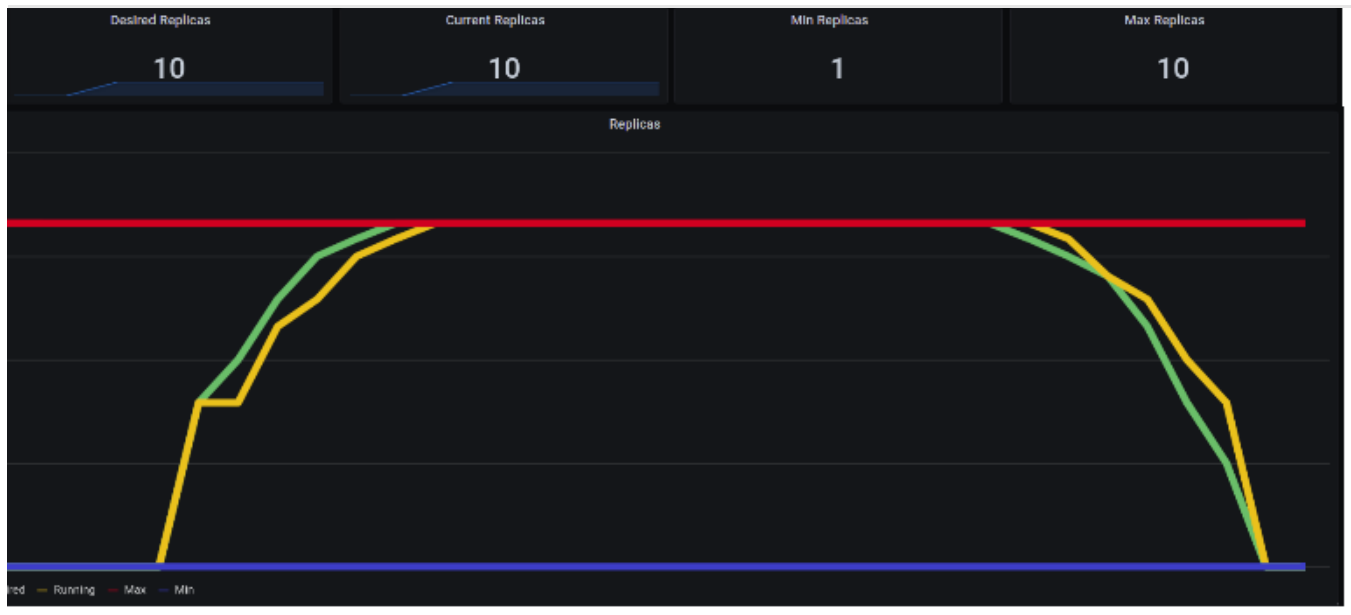

Figure 14. Replicas creation and deletion.

We have a traffic generator in our scenario. When Locust start generating traffics, we are able to see charts created in their web-interface. In the following figures, charts created by Locust such as the number of users, video sink´s pods response times and total requests per second for a short testing interval will be illustrated to assess the validity of test.

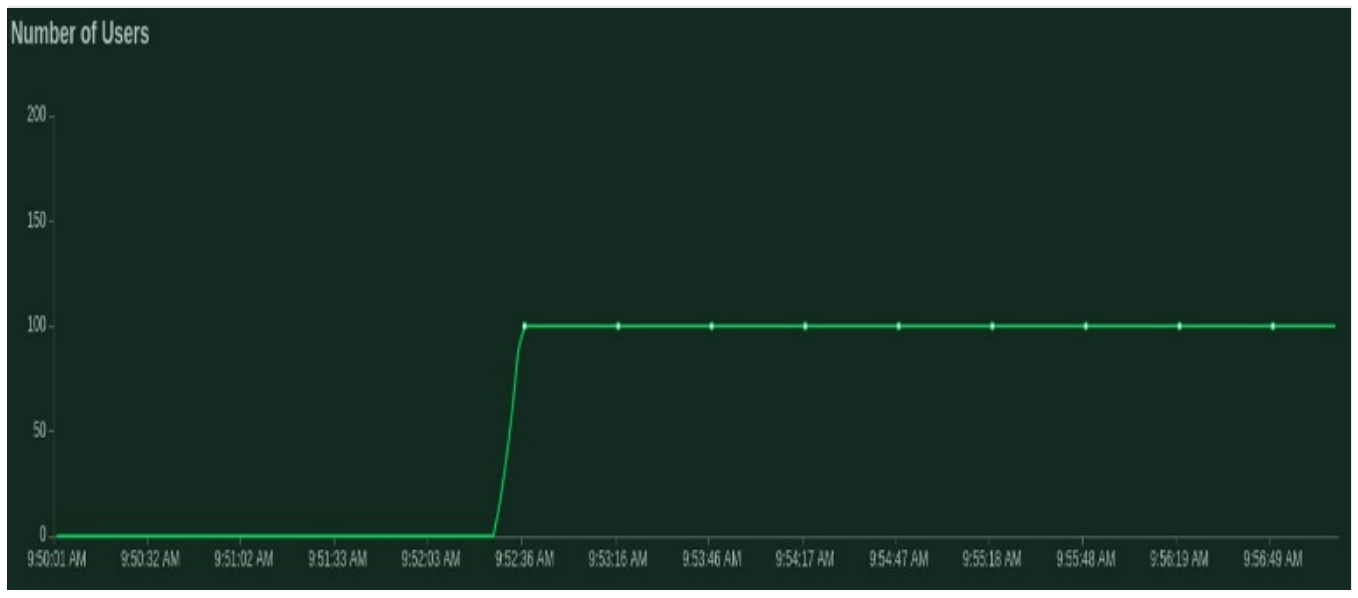

Figure 15. Number of users (100).

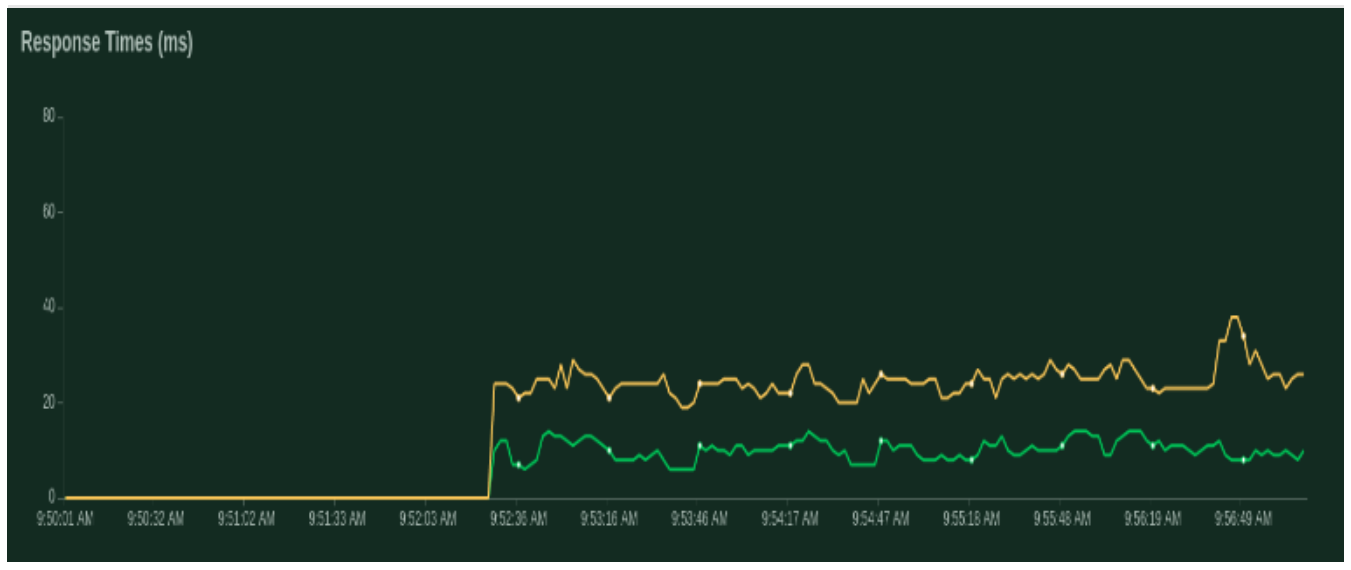

Figure 16. Video Sink’s pods Response time.

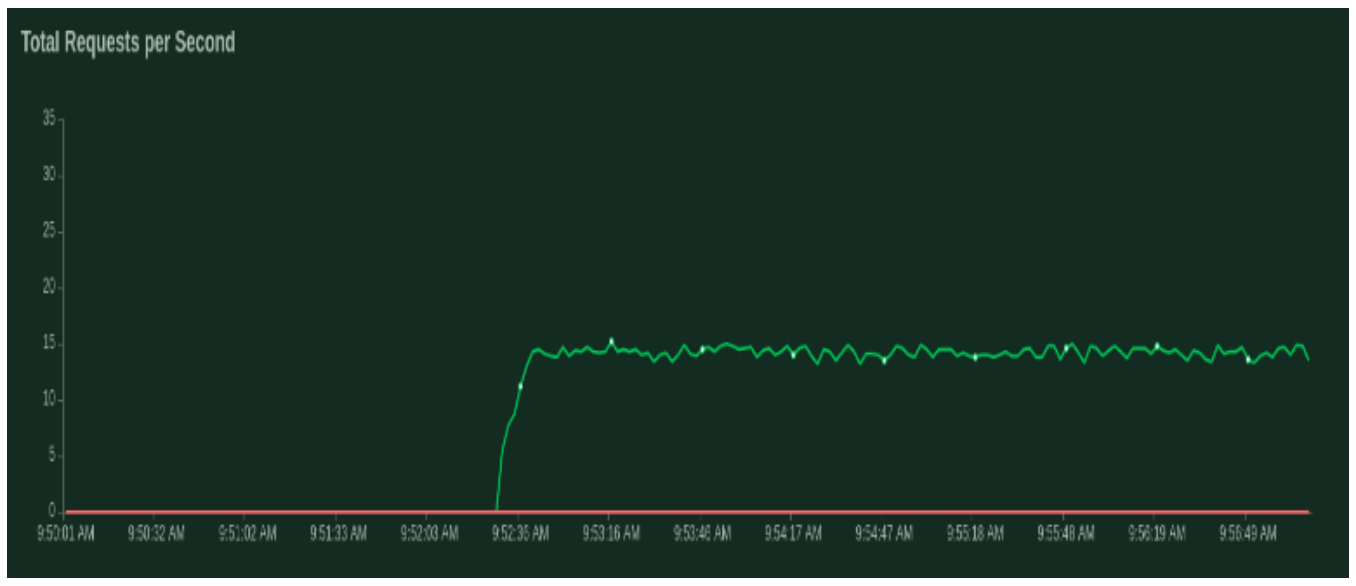

Figure 17. Total number of requests per second.

## V. CONCLUSION

In this article, we illustrate our 5G scalable testbed architecture design. The testbed is designed and implemented to be scaled in different instances for different scenarios accordingly. We test an emulated multimedia scenario in our infrastructure. The deployment of multimedia scenario is in our Kubernetes infrastructure connected with Open5GS core and Amariosoft where we were able to scale in/out emulated video sinks according to service metric (HTTP Requests). In our future work, we want to test and show that testbed instances scalability can be effective 5G network key performance indicator (KPIs).

## ACKNOWLEDGMENT

This work is founded by the H2020 5GSolutions (Grant Agreement no.856691) and H202 5GMediaHUB (Grant Agreement no.101016714).